# Spectral signature of atmospheric winds in high resolution transit observations

Engin Keles

*Leibniz-Institut für Astrophysik Potsdam (AIP), An der Sternwarte 16, 14482 Potsdam, Germany*

**ABSTRACT**

The study of exoplanet atmospheres showed large diversity compared to the planets in our solar system. Especially Jupiter type exoplanets orbiting their host star in close orbits, the so-called hot and ultra-hot Jupiters, have been studied in detail due to their enhanced atmospheric signature. Due to their tidally locked status, the temperature difference between the day- and nightside triggers atmospheric winds which can lead to various fingerprints in the observations. Spatially resolved absorption lines during transit such as sodium (Na) could be a good tracer for such winds. Different works resolved the Na- absorption lines on different exoplanets which show different line widths. Assuming that this could be attributed to such zonal jet streams, this work models the effect of such winds on synthetic absorption lines. For this, transiting Jupiter type planets with rotational velocities similar to hot and ultra-hot Jupiter are considered. The investigation shows that high wind velocities could reproduce the broadening of Na-line profiles inferred in different high-resolution transit observations. There is a tendency that the broadening values decrease for planets with lower equilibrium temperature. This could be explained by atmospheric drag induced by the ionization of alkali lines which slow down the zonal jet streams, favoring their existence on hot Jupiter rather than ultra-hot Jupiter.

## 1 INTRODUCTION

The last decades showed the existence of Jupiter like giant planets in very close orbits to their host stars. Although being comparable in size and mass with Jupiter, the extrasolar hot and ultra hot Jupiter type planets are exposed by strong stellar radiation, leading to much different atmospheric dynamics (and therefore atmospheric fingerprints) compared to Jupiter in our solar systems. Due to the expected tidally locked status of these planets where the rotation and orbital periods become equal, the planetary hemispheres separate into a permanent night- and a dayside making these atmospheres more turbulent compared to Jupiter with temperature contrasts of 500K - 1000K (Knutson et al. 2008) depending on the vertical atmospheric advection and radiation timescales (Heng & Showman 2015). The energy budget of the planetary atmosphere becomes dominated by the intense radiation, where the differential heating of the atmosphere induces pressure gradient forces resulting in atmospheric motions i.e. winds (Dobbs-Dixon et al. (2010); Perna et al. (2010); Heng & Showman (2015)). The wind speeds are a result of a balance between the atmospheric kinetic energy and the sustained friction within the atmosphere. Coming across, although tidally locked, the planetary rotational velocities can still be large especially for ultra hot Jupiter (UHJ) type planets (e.g. ∼6.6 km/s for KELT-9b or ∼7 km/s for WASP-33b) leading to the combination of both. The redistribution of energy in planetary atmospheres triggering the atmospheric circulations is a complex issue where the radiative transfer and atmospheric dynamics is addressed in different 3D modeling approaches see e.g. Showman et al. (2008), Dobbs-Dixon et al. (2010), Lian & Showman (2010), Perna et al. (2010), Showman & Polvani (2011), Miller-Ricci Kempton & Rauscher (2012), Showman et al. (2012), Kempton et al. (2014), Zhang et al. (2017), Flowers et al. (2019), Steinrueck et al. (2019) or Debrecht et al. (2020). For hot Jupiter (HJ) type planets, these studies show two main outcomes related with atmospheric winds, being an equatorial super-rotating jet around 1 bar pressure moving into the direction of planetary rotation with wind velocities around a few km/s (Showman et al. 2008; Showman & Polvani 2011; Showman et al. 2012) and the existence of day-to-nightside winds with velocities on the order of 1 - 10 km/s at mbar pressure regime (Miller-Ricci Kempton & Rauscher 2012). In the case of the day-to-nightside winds, the winds become stronger for larger day-to-nightside temperature contrasts, pushing the atmosphere into the nightside direction (Showman et al. 2012). Note, that the wind velocities and directions depend strongly on the frictional drag within the atmosphere (see for instance Figure 3 in Miller-Ricci Kempton & Rauscher (2012)) and can show timely variable conditions (Komacek & Showman 2019). For instance, one kind of friction is induced by the planetary magnetic field which tends to reduce wind velocities of weakly ionized winds by magnetic drag, as the thermally ionized particles interact with this field (Perna et al. 2010).

That atmospheric dynamics can cause signals in transmission spectra by acting on the absorption profiles from the planetary atmosphere is known for decades and comprehensively discussed in detail in e.g. Brown (2001), but also in combination with high resolution observations in very recent studies such as Seidel et al. (2020) and Cauley et al. (2020). Several works investigated and/or showed wind properties for transiting exoplanets using observations and modeling approaches e.g. Spiegel et al. (2007), Snellen et al. (2010), Miller-Ricci Kempton & Rauscher (2012), Showman et al. (2012), Kempton et al. (2014), Wyttenbach et al. (2015), Louden & Wheatley (2015), Brogi et al. (2016), Flowers et al. (2019), Seidel et al. (2019), Casasayas-Barris et al. (2019), Cauley et al. (2019), Seidel et al. (2020), Gebek & Oza (2020), Keles et al. (2020), Ehrenreich et al. (2020), Cauley et al. (2020) and more. Two categories seem to be of main interest for the majority of the studies investigating atomic and molecular lines: On the one side, the investigation of day-to-nightside winds by searching for the blueshift of spectral lines, where it is expected that most irradiated planets should show the strongest blueshifts (Miller-Ricci Kempton & Rauscher 2012; Showman et al. 2012). On the other side, the effect of atmospheric circulation on spectral lines, mainly sodium (Na) for the atomic lines and carbon monoxide for the molecular lines, which result in line deformation mostly due to line broadening.

One reason to use the Na- absorption lines as tracer for atmospheric circulation is due to its strong atmospheric opacity on HJ type planets leading to strong absorption signature. In recent years, several authors resolved Na lines on exoplanets in high resolution observations e.g. on HD189733b (Wyttenbach et al. 2015), on WASP-69b (Casasayas-Barris et al. 2017), on WASP-49b (Wyttenbach et al. 2017), on KELT-20b (Casasayas-Barris et al. 2019), on WASP-17b (Khalafinejad et al. 2018), on WASP-76b (Seidel et al. 2019), on WASP-127b (Žák et al. 2019) and on WASP-52b (Chen et al. 2020), whereas in several cases the Na-lines show blueshifts due to day-to-nightside winds and very large full width at half maximum (FWHM) values, the latter hinting that some mechanism seems to broaden these lines.

This work aims to investigate the effect of super-rotating atmospheres with different wind patterns on Na absorption line profiles on Jupiter type planets inferred during a planetary transit. The most comparable study to this work is shown by Spiegel et al. (2007), who investigated the effect of planetary rotation on the Na-lines in transmission spectra for the HJ HD209458b, giving a nice overview of the different dynamical stellar and planetary effects affecting the transmission spectra. However, this work concentrates only on planetary rotation neglecting other kinds of atmospheric motion in contrast to that is presented in this work. Note, that although Na-D2-line deformation is investigated, the effect is similar also for other spectral lines which can be resolved in planetary atmospheres such as magnesium (see e.g. Cauley et al. (2019)), calcium (see e.g. Yan et al. (2019)), iron (see e.g. Casasayas-Barris et al. (2019)), potassium (see e.g. Chen et al. (2020)) and more. For lighter elements probing very high altitudes resolved in planetary atmospheres such as Hydrogen (see e.g. Jensen et al. (2018)) or Helium (see e.g. Nortmann et al. (2018)) evaporative winds may play another significant role (Salz et al. 2016, 2018) as well as line broadening further induced by large number densities which lead to optically thick absorption layers (Huang et al. 2017; Wyttenbach et al. 2020).

This paper is structured as follows. In Section 2 the model used is described. Section 3 explains the methodology of this paper. The results are presented in Section 4. Section 5 discuss the results and caveats of the application and Section 6 presents the summary.

## 2 MODEL

The effect of a rotating atmosphere on synthetic absorption lines is modelled using a grid model that maps the planetary surface by 200 x 200 pixels assuming that the planet's spin axis is perpendicular to its orbital plane which is in line-of-sight. Each pixel on the spherically symmetric atmosphere contains the atmospheric Na-D2-absorption line profile from Keles et al. (2020) for the exoplanet HD189733b to make the investigation more applicable to observational findings. The authors compared synthetic transmission spectra with the resolved Na- lines observed by Casasayas-Barris et al. (2017) for this exoplanet and derived the unbroadened Na-D2-line profile (see Gaidos et al. (2017); Bower et al. (2019); Keles et al. (2020) for further details on the synthetic line modelling approach). The right panel in Figure 1 shows the Na-D2 line profile (yellow). The model assumes rigid body rotation and each spectral line in each pixel is Doppler-shifted according to its distance to the planetary rotation axis. Assuming a constant angular velocity of the atmosphere, the line of sight velocity is calculated via $V_{LOS} = V_{rot} \times \sin\theta$ with $V_{rot}$ being the planetary rotation velocity and $\theta$ being the line of sight angle. The absorption arising from the atmosphere with the largest distance to the rotation axis ($\theta = 90°$) shows the largest Doppler shift. Additional winds are dealt additive in each pixel e.g. in case of zonal jet streams, the line of sight velocity is the sum of all velocity contributions i.e. $V_{LOS} = (V_{rot} + V_{Jet}) \times \sin\theta$. Day-to-nightside winds are not included in the model, assuming that these act in the same way at all latitudes and longitudes, leading to an overall net blueshift of the final line profiles without inducing a line deformation.

Figure 1 demonstrates the model. The left panel shows a Jupiter type planet with $V_{rot}$ = 6 km/s, where only the atmospheric circle at $R_{pl} < R_{Atmo} < 1.3 \times R_{pl}$ (with $R_{pl}$ being the white light radius of the planet) contributes to the transmission signal. The middle panel shows few Doppler-shifted absorption profiles (dashed) arising from the different parts of the atmosphere for illustration purposes, whereas the final line profile (solid black) is the sum over all pixels on the atmospheric circle. The right panel shows the line profile before broadening (yellow) and the line profile after broadening (black). For demonstration purposes, the dashed lines show the broadened line profiles not accounting only for the atmospheric circle but the full surface (i.e. similar to stellar surfaces) from the model used in this work (magenta) and the [1] PyAstronomy tool *pyasl.fastRotBroad* (green) which introduces rotational broadening for stellar surfaces according to (Gray & Corbally 1994). Accounting only for absorption by a transparent circle leads to a broader and shallower line profile. Therefore, it is important to account for this effect to not over-/underestimate the velocity amplitudes needed to broaden spectral lines, especially for large velocities as the deviation increases non-linear with increasing velocity amplitude.

## 3 METHOD

This study aims to investigate the line deformation of spectral lines induced by atmospheric winds in form of zonal jet streams in high resolution transmission spectra. For this, it is assumed that the atmospheric motion introduces a Doppler-shift to the spectral lines leading to a velocity broadening in form of rotational broadening, thus not changing the intrinsic Doppler-width of the absorption lines from each pixel element from the transparent atmospheric circle. Shown is the integrated absorption line profile in the planetary rest frame, assuming that the high spectral resolution spectra taken at short exposure times (to avoid Doppler smearing, see e.g. Cauley et al. (2020)) are corrected for spurious observational and instrumental effects (see e.g. Wyttenbach et al. (2020) for a comprehensive overview on extracting transmission spectra). Such effects could be telluric contamination, stellar activity or velocity shifts due to the planetary orbital motion, whereby in the latter case one has to consider non-zero eccentricities, non-edge orbits, or other factors that could lead to radial velocity shifts in the residual spectra, which could be misunderstood as the effect of different kind of winds (Zhang et al. 2017).

---

[1] https://github.com/sczesla/PyAstronomy

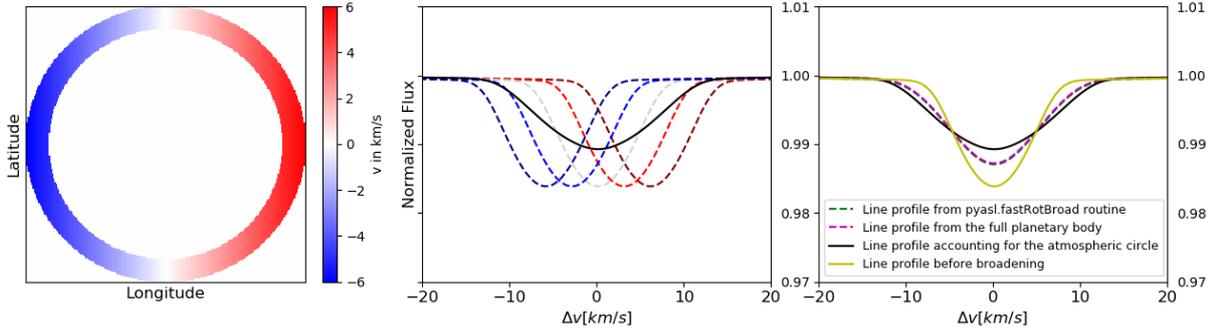

**Figure 1.** Effect of rotational broadening on absorption lines. Left panel: Illustration of an absorbing circle from the transparent region of a Jupiter type planet with a planetary rotation of ± 6 km/s. Middle panel: Dashed lines show absorption lines that are shifted in velocity with the colour coding illustrating the different regions from the atmospheric circle. The black solid line shows the final line profile which is the sum of all velocity shifted contributions. Note, that there is an overdensity of line profiles around the line core region of the black solid line which are not shown for better visibility of the illustration. Right panel: Yellow solid line shows the line profile before broadening and the black solid line after the broadening. The dashed lines show the broadening if one accounts for the full surface i.e. similar to a stellar surface and compares this work (magenta) to the PyAstronomy tool *pyasl.fastRotBroad* (green).

Different scenarios will be investigated showing different wind patterns for a Jupiter type planet with $V_{rot}$ = 3 km/s and with $V_{rot}$ = 6 km/s. The rotational velocity of $V_{rot}$ = 3 km/s is representative for the rotational velocity of an HJ (see e.g. HD189733b with ∼2.7 km/s) and the rotational velocity of $V_{rot}$ = 6 km/s representative for the rotational velocity of an UHJ (see e.g. KELT-9b with ∼6.6 km/s). The scenarios presented are described below (see corresponding sections for further details on the motivation):

(i) The rotational broadening in an extended ($R_{pl} < R_{Atmo} < 1.1 \times R_{pl}$) and unextended ($R_{pl} < R_{Atmo} < 1.5 \times R_{pl}$) atmosphere with a weaker and stronger Na- absorption line (Section 4.1)

(ii) A eastward streaming equatorial zonal jet with $V_{Jet}$ = ∓ 6 km/s at the trailing and leading limb respectively and different widths reaching latitudes of ±20° and ±40° (Section 4.2)

(iii) A eastward streaming equatorial zonal jet with asymmetric velocities on the limbs being $V_{Jet}$ = ∓6 km/s at the trailing and $V_{Jet}$ = ∓2 km/s at the leading limb (i.e. a third velocity amplitude) reaching latitudes of ±20° and ±40° (Section 4.3)

(iv) A eastward streaming equatorial zonal jet with $V_{Jet}$ = ∓ 6 km/s at the trailing and leading limb respectively reaching a latitude of ±20° and westward streaming zonal jets at the poles with $V_{WJet}$ = ± 2 km/s and $V_{WJet}$ = ± 6 km/s (Section 4.4)

(v) Timely resolved spectra during ingress and egress for the demonstrated science cases in Section 4.2, Section 4.3 and Section 4.4 showing a Jupiter type planet with $V_{rot}$ = 3 km/s (Section 4.5)

To investigate if the wind structure according to the Scenarios (ii)-(v) can be differentiated from the profiles that are not accounting for winds or even between the different wind properties, the required observational stellar spectral signal-to-noise (S/N) values are shown in the corresponding sections (i.e. Section 4.2, Section 4.3, Section 4.4 and Section 4.5). The quantitative determination of a wind pattern with high significance will require high spectrally and timely resolved observations with high S/N ratios by observations with current high-resolution spectrographs. The required spectral mean S/N from the stellar spectrum can be approximated by:

$$S/N_{n\sigma} \approx \frac{1}{|\overline{\Delta F}|} \times \frac{n\sigma}{\sqrt{m} \times \sqrt{obs} \times \sqrt{spec} \times \sqrt{lines}} \approx \frac{1}{|\overline{\Delta F}|} \times \Phi_{Planet} \quad (1)$$

Here, "$|\overline{\Delta F}|$" is the mean of the absolute flux difference between the modelled line profiles i.e. the difference between the broadened and unbroadened line profiles as well as the difference between the broadened line profiles with different wind properties. "$(|\overline{\Delta F}|)^{-1}$" is approximated as the required mean S/N per resolved spectral resolution element (neglecting minor error contributions such as from the "Master-out" spectrum, which is used to infer the transmission profile, see e.g. Wyttenbach et al. (2020)). In the second term, abbreviated by $\Phi_{Planet}$, $n\sigma$ is the significance of the detection, "m" the number of resolved points, "obs" the number of observed transits, "spec" the number of snapshots for the same pattern for one transit (i.e. one for ingress/ egress observations but several during the in-transit phase) and "lines" the number of spectral lines which can be combined (assuming similar absorption line strengths). To investigate the detectability of the wind pattern on a real test case, $\Phi_{Planet}$ is determined for the exoplanet HD189733b for a $3\sigma$ ($n\sigma$ = 3) detection and one transit (obs = 1), with m = 58 (corresponding to points ±15 km/s around the line core), lines = 2 (assuming that 2 Na-D-lines are used) and spec ≈ 18 (assuming that an exposure time of 4 min (neglecting readout time) is taken between the second and third contact of HD189733b). This results in $\Phi_{HD189733b}$ ≈ 0.066 during the in-transit phase Scenarios (ii)-(iv) and $\Phi_{HD189733b}$ ≈ 0.279 for Scenario (v), because for the latter, only single snapshots are possible.

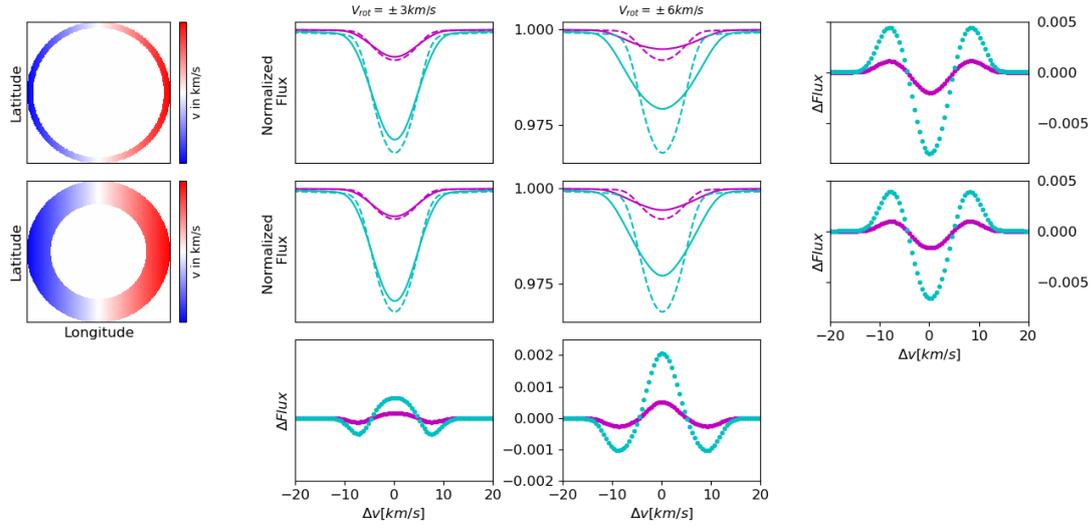

**Figure 2.** The effect of rotational broadening on a stronger (magenta) and weaker (purple) Na absorption line before (dashed) and after (solid) broadening. The Na-D2-line from Keles et al. (2020) is scaled for this purpose. The translucent atmospheric circle is set to ($R_{pl} < R_{Atmo} < 1.1 \times R_{pl}$) (first line) and ($R_{pl} < R_{Atmo} < 1.5 \times R_{pl}$) (second line). The second column shows a Jupiter type planet with $V_{rot} = 3$ km/s and the third with $V_{rot} = 6$ km/s. The first column illustrates the science cases. The third row shows the difference between the broadened line profiles from the panels above and the forth column shows the difference between the broadened line profile from the different rotating planets (each for the same line strength). Each column shares the x-axis and the second and third column share the same y-axis.

The computational time of the different wind structure in this work is cheap requiring only a few seconds each on a laptop computer with Intel Core i7 -10510U CPU. But note, that 1D modeling procedures may underestimate physical effects of particular importance considered in 3D modelling approaches (Kempton et al. 2014). The interplay between atmospheric dynamics and rotation is a complex issue and the atmospheric Doppler signature a combination of multidimensional wind fields and rotation rate (Flowers et al. 2019). Also other effects such as high-altitude clouds, atmospheric metallicities and atmospheric temperature inversion can affect the heat redistribution in those atmospheres (Zhang et al. 2018), making the application of advanced general circulation modells necessary. Several caveats are discussed in Section 5.

## 4 RESULTS

### 4.1 Rotational broadening of weaker and stronger Na- lines in an extended and unextended atmosphere

Exoplanet atmospheres of Jupiter type planets can show big diversity e.g. in gravity, temperature and Na-abundance. The atmospheres can have small and large scale heights with weaker and stronger absorption lines, making the fingerprints of these rotating atmospheres diverse in transmission spectra.

To account for such diversities, Figure 2 shows a Jupiter type planet with $V_{rot} = 3$ km/s and $V_{rot} = 6$ km/s with an extened and unextended atmosphere showing a weak and a strong Na-absorption line (see Figure caption for further information). Inspecting the first row with the unextended atmosphere, the rotating atmospheres show broadened line profiles for both $V_{rot}$ cases where the weaker and stronger line profiles become shallower and broader compared to the non-broadened ones as expected. This effect is stronger for the faster rotating planet due to the stronger Doppler-shift of the line profiles and significantly stronger visible for the stronger line profiles compared to the weaker ones in sense of absolute flux values, also suggested by the difference plot in the fourth column.

Inspecting the second row with an extended atmosphere, the line deformations look quite similar to that is demonstrated for the unextended atmosphere. Comparing the difference in line profiles for an extended and unextended atmosphere, hence the panels in the third row, the difference is small for the slower rotating planet and increases for the faster rotating one, again the effect being stronger for the stronger absorption line profiles. The positive peak at the line core region shows that the line profiles from the extended atmosphere are deeper, thus showing less broadening, compared to the unextended atmospheres. This is due to the enhanced contribution to the line core region from Na-lines which are less shifted in velocity arising from the wider transparent atmospheric circle. Thus, assuming that the rotational velocity is the same on the trailing and leading limb, the contribution of absorption lines which are less shifted in velocity is larger for unextended atmospheres than for extended, as the line of sight contribution decreases with longitude.

In conclusion, line broadening (or deformation) is stronger for faster rotating planets, whereas the width of the translucent atmospheric ring is of relatively less importance compared if one has in mind that the extended atmosphere is four times wider in our scenario (note, that the line strength is assumed to be the same varying $R_{atmo}$ in the model used, so that the extension only affects the contribution of different Doppler-shifts). For large rotational velocities, weak absorption lines can become very shallow (e.g. see third column) and could be hidden within the noise level of transmission spectra. However, note that the equivalent width of the lines is conserved in case of rotational broadening. Due to this, the absorption signature may be inferred with the so called "excess method" where the flux is integrated within passbands and normalized thereafter showing the atmospheric absorption in excess light curves (see e.g. Keles et al. (2019)). The atmospheric extension will be fixed to ($R_{pl} < R_{Atmo} < 1.3 \times R_{pl}$) for the upcoming scenarios and not varied, as the variation has a negligible effect on the qualitative outcome of the results.

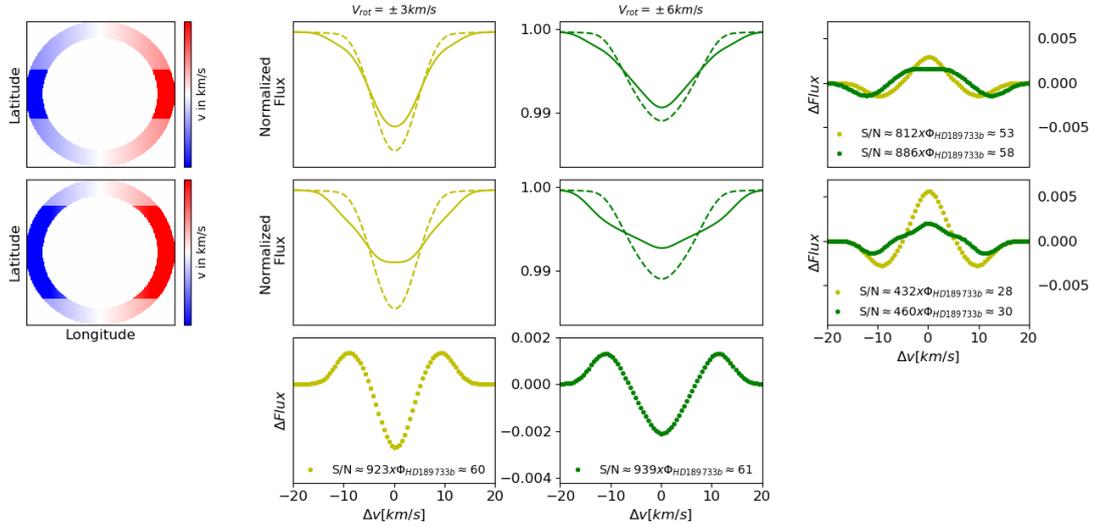

**Figure 3.** The effect of an eastward streaming jet with $V_{Jet} = \mp 6$ km/s $\times \sin\theta$ reaching latitudes of $\pm 20°$ (first row) and $\pm 40°$ (second row) on Na-absorption lines (solid) on a planet with $V_{rot} = 3$ km/s (second column) and $V_{rot} = 6$ km/s (third column). The dashed lines consider only the planetary rotation for comparison. The atmospheric extension is fixed to ($R_{pl} < R_{Atmo} < 1.3 \times R_{pl}$). The first column illustrates the science cases. The fourth column show the difference between the broadened and unbroadened line profiles and the third row show the difference for the broadened line profiles with different jet streams. The label shows the $3\sigma$ spectral mean S/N value with $\Phi_{HD189733b} \approx 0.066$. Each column shares the x-axis and the second and third column share the same y-axis.

### 4.2 Effect of an eastward streaming zonal jet reaching latitudes of $\pm 20°$ and $\pm 40°$ on the Na- absorption line

This scenario investigates the effect of an zonal jet stream on a Jupiter type planet with $V_{rot} = \pm 3$ km/s and $V_{rot} = \pm 6$ km/s. Such a jet stream can extend from the equator to latitudes around 20°-60°, thus dominating the surface of such a giant planet (Showman & Polvani 2011; Showman et al. 2012). The width of such a jet stream is controlled by the Rossby deformation radius, which increases for larger atmospheric scale height and slower planetary rotation rates (Showman et al. 2008; Heng & Showman 2015). The existence of such jet streams on a planet depend on the one side on the stellar irradiation and so the radiative time constant and on the other side on the frictional drag within the atmosphere (Heng & Showman 2015). For strongly irradiated planets with weak atmospheric frictional drag, day-to-nightside winds at low pressure level dominate but also zonal jet streams at higher pressure level can originate. Decreasing the stellar insolation, zonal jet streams dominates at lower altitudes, which become unable to form increasing the strength of the frictional drag within the atmosphere (Showman et al. 2012). Thus, planets can posses day-to-nightside winds, zonal jet streams or even both. Note, that the jet streams here are assumed to be zonally present, but magnetic fields can also perturb the circulation in ionized hot Jupiter atmospheres leading to circulation pattern not being anymore zonal (see e.g. Figure 3 in Batygin & Stanley (2014)). Moreover, atmospheric variability due to magnetohydrodynamic as well as hydrodynamic variability can lead to variable conditions on hot Jupiter atmospheres such as reversing the direction of the equatorial jet or shifting these in latitude (Komacek & Showman 2019).

Figure 3 shows the effect of an eastward zonal jet stream reaching latitudes of $\pm 20°$ and $\pm 40°$ with a $V_{jet} = \mp 6$ km/s $\times \sin\theta$ (see Figure caption for further information). Note, that the velocities are dealt additive, thus the maximum velocity at the trailing and leading limb is $V_{LOS} = \mp V_{rot} + \mp V_{jet}$, decreasing with $\sim \sin\theta$. Investigating the first row, the Na- absorption line profiles become broadened similar to that is presented in Section 4.1, but are deformed more strongly. For the slower rotating planet, the zonal jet stream lead to deformations in the line core (which become flatter) and line wings (which become broader) due to the contribution of the Na-lines which are stronger Doppler-shifted by the zonal jet stream. In comparison, for the faster rotating planet, the effect is less pronounced especially in the line core region. This is not surprising, as the combined maximum line of sight velocity is three times larger than the planetary rotation for the slower rotator, but only two times larger for the faster rotator, affecting stronger the line profiles from slower rotating planetary atmosphere. This is also suggested from the flux differences shown in the fourth column.

Investigating the second row where the zonal jet stream reaches higher latitudes, the difference increases for both cases (also suggested by the difference plots in the third row), as the contribution from the Doppler-shifted Na- lines from the polar regions (arising from the planetary rotation) decreases. For both cases, the line cores flatten even more, the reason being the same as mentioned above.

The required spectral S/N values show that wider jet streams are easier to detect than narrower ones. The differentiation if a jet stream reaches either $\pm 20°$ or $\pm 40°$ is similarly difficult for both types of rotating planets.

In conclusion, the main difference between a streaming zonal jet reaching latitudes of $\pm 20°$ and $\pm 40°$ is the missing contribution of Doppler-shifted absorption lines arising from the planetary rotation at the line core region, but the enhanced contribution of Doppler-shifted absorption lines arising from the zonal jet stream at the region where the line wings arise. This leads the lines becoming shallower and broader with increasing the width of the streaming jet. In case of unextended atmospheres and large Doppler-shifts, the missing contribution from low velocity shifted absorption lines may introduce a w-shaped feature at the line core region (not shown here), which decreases with increasing contribution from less Doppler-shifted absorption lines. Wide zonal jet streams will be easier to detect on slower rotating Jupiter type planets compared to faster rotating ones as the line deformation show a larger contrast compared to the line broadening induced by planetary rotation.

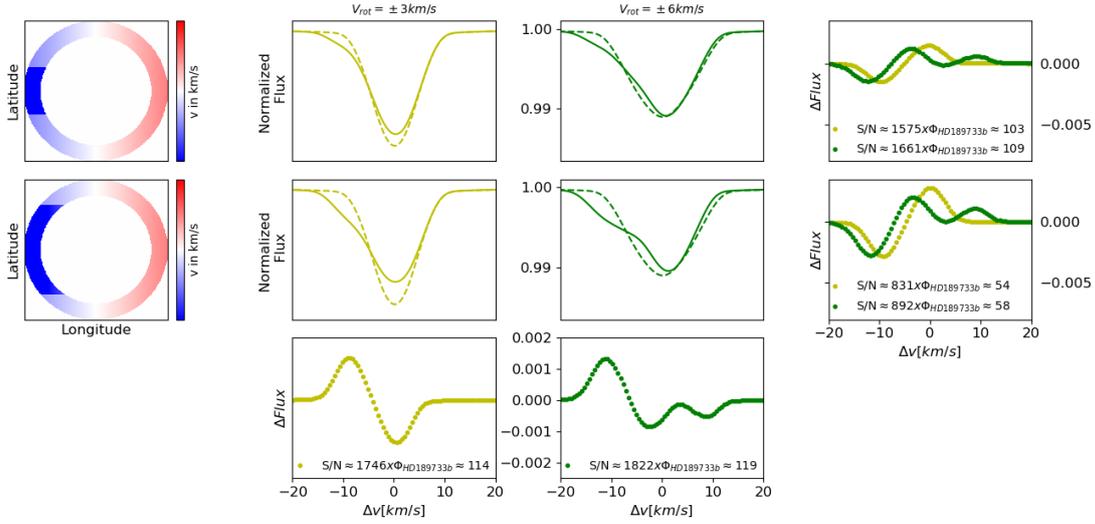

**Figure 4.** The effect of an eastward streaming jet with with $V_{JET} = \mp 6$ km/s and line of sight velocity of $V_{E-LOS} = (V_{rot} + V_{JET}) \times \sin\theta$ on the eastern hemisphere and $V_{W-LOS} = 0.33 \times (V_{rot} + V_{JET}) \times \sin\theta$ on the western hemisphere reaching latitudes of $\pm 20°$ (first row) and $\pm 40°$ (second row) on Na-absorption lines (solid) on a planet with $V_{rot} = 3$ km/s (second column) and $V_{rot} = 6$ km/s (third column). The dashed lines consider only the planetary rotation for comparison. The atmospheric extension is fixed to ($R_{pl} < R_{Atmo} < 1.3 \times R_{pl}$). The first column illustrates the science cases. The fourth column show the difference between the broadened and unbroadened line profiles and the third row show the difference for the broadened line profiles with different jet streams. The label shows the $3\sigma$ spectral mean S/N value with $\Phi_{HD189733b} \approx 0.066$. Each column shares the x-axis and the second and third column share the y-axis.

### 4.3 Effect of an eastward streaming zonal jet with different velocities on the trailing and leading limb reaching latitudes of ±20° and ±40° on the Na- absorption line

Due to tidal locking of close-in planets, the substellar point on the planetary dayside becomes the hottest region, the so called "hot spot". Equatorial streaming zonal jets can shift this "hot spot" into the direction of the planetary rotation (Miller-Ricci Kempton & Rauscher 2012). This "hot spot" can have a large influence on the atmospheric wind pattern, leading to an asymmetric velocity pattern (see e.g. Figure 4 in Miller-Ricci Kempton & Rauscher (2012) or Figure 6, 7 and 8 in Showman et al. (2012)). One example for such a influence could be a shifted "hot spot" which triggers day-to-nightside winds with different strengths on the atmospheric transparent circle, which combine with the planetary rotation and streaming zonal jet to asymmetric wind pattern on the both hemispheres. Furthermore, asymmetric wind pattern could be also arise if the planets rotational axis is not perpendicular to the orbital plane (Cauley et al. 2020). Non-uniform wind patterns within the atmosphere can lead to asymmetric Doppler-shifts of the Na- absorption lines on the translucent atmospheric ring resulting in asymmetric absorption line profiles (Kempton et al. 2014; Flowers et al. 2019).

To investigate on such a scenario, Figure 4 shows the same wind pattern as shown in Section 4.2 but with different line of sight velocities on the eastern and western hemispheres (i.e. $V_{E-LOS} = (V_{rot} + V_{JET}) \times \sin\theta$ on the eastern hemisphere and $V_{W-LOS} = 0.33 \times (V_{rot} + V_{JET}) \times \sin\theta$ on the western hemisphere, thus a third in velocity amplitude).

Investigating the first row, the line profiles become highly asymmetric compared to the line profiles shown in Section 4.2, where the absolute streaming jet velocity is the same on both hemispheres. This is explained by the difference in Doppler-shift from both hemispheres, where at the western hemisphere the Na- absorption lines are stronger blueshifted than redshifted on the eastern hemisphere. Due to this difference in Doppler-shifts, the line profile is not anymore symmetrical with respect to the line center. For the slower rotating planet, the maximum line of sight velocity at the leading limb becomes $V_{LOS} = 3$ km/s and thus the same to $V_{rot}$ so that the line wings overlap with the dashed line profile which considers only the planetary rotation. For the faster rotating planet the maximum line of sight velocity becomes $V_{LOS} = 4$ km/s and thus 50% smaller than the planetary rotation velocity, leading to a slight difference in the line wings. However, at the trailing limb, the line profiles are more strongly deformed similar to that is shown in Section 4.2, with the difference that the line cores are deeper in this scenario due to the larger contribution from the lower velocity Doppler-shifted Na- absorption lines from the eastern hemisphere.

Increasing the width of the streaming jet (second row), the effects become stronger leading to stronger line deformation, especially blueward the line core due to the increase in Na- absorption lines which are Doppler-shifted for a wider jet stream whereas the difference is again small redward the line core, the reason being the same as mentioned above. The panel in the third column and fourth row showing the flux differences illustrate the mentioned deviations in the line profiles.

Similar to Scenario 4.2, the wider jet streams are easier to detect and the differentiation of the jet stream width is similarly difficult for both type of rotating Jupiter type planets, with the difference that the required spectral S/N values are around a factor of two higher compared to the wind pattern without asymmetric velocities on the hemispheres.

In conclusion, the missing high velocity contribution at the eastern hemisphere has two main effects on the line profiles: First, the line profile becomes more asymmetric due to the difference in blueshifted and redshifted absorption lines from the eastern and western hemisphere, which increases with increasing width of the zonal jet stream. Second, as line properties of resolved line profiles are usually derived by Gaussian fits assuming symmetric lines, the asymmetric Doppler-shifted line profiles would mimic a blueshift of the line core. Thus asymmetric wind patterns can mimic blueshifts of resolved line profiles especially for planets with lower $V_{rot}$ and strong asymmetric velocities on the two planetary hemispheres.

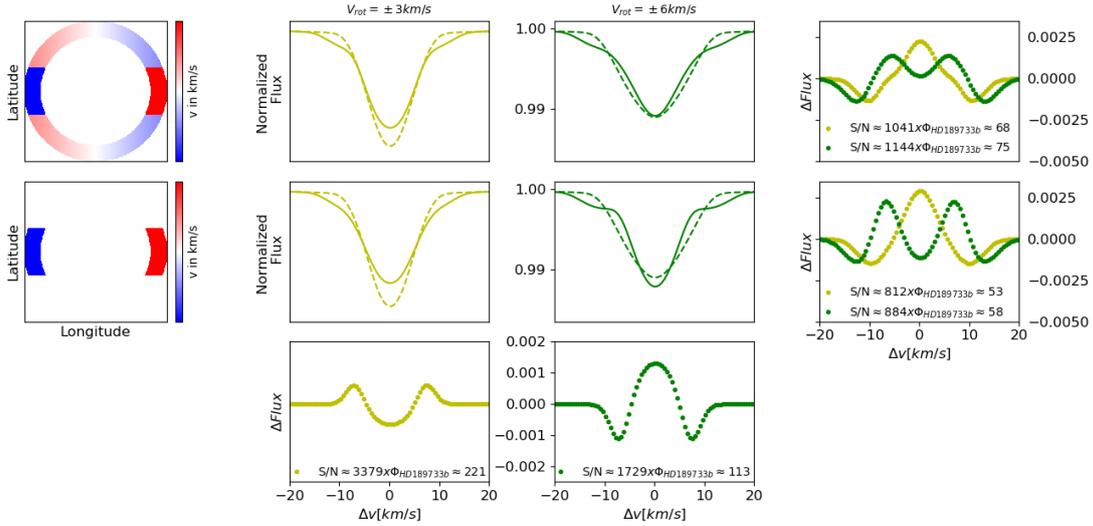

**Figure 5.** The effect of an eastward streaming jet with $V_{Jet} = \mp 6$ km/s reaching latitudes of $\pm 20°$ and two westward streaming jets at the poles with $V_{WJET} = \pm 2$ km/s (first row) and $V_{WJET} = \pm 6$ km/s (second row) on Na-absorption lines (solid) on a planet with $V_{rot} = 3$ km/s (second column) and $V_{rot} = 6$ km/s (third column). The dashed lines consider only the planetary rotation for comparison. The atmospheric extension is fixed to ($R_{pl} < R_{Atmo} < 1.3 \times R_{pl}$). The first column illustrates the science cases for the faster rotating planet. The fourth column show the difference between the broadened and unbroadened line profiles and the third row show the difference for the broadened line profiles with different westward streaming jet velocities. The label shows the $3\sigma$ spectral mean S/N value with $\Phi_{HD189733b} \approx 0.066$. Each column shares the x-axis and the second and third column share the same y-axis.

### 4.4 Effect of an eastward streaming zonal jet reaching latitudes of ±20° with additional westward streaming jets at the poles on the Na- absorption line

In the solar system, the gas giants show a wide equatorial eastward streaming jet and additionally large number of smaller zonal jets streaming from east to west, whereas the ice giants have a small number of eastward streaming jets at high latitudes additional to a sub-rotating equatorial westward streaming jet. A explanation for the differences is shown in Lian & Showman (2010), where the water abundance (which is way different for the ice giants and gas giants) and the corresponding large scale latent heating of the atmosphere due to the condensation of water vapor seems to play a major role. For Jupiter type exoplanets, westward streaming jets with lower velocity amplitudes, which are triggered from the hot dayside, can also arise additional to an eastward streaming equatorial jet (see e.g. Showman et al. (2008)).

Figure 5 investigates a scenario on a Jupiter type planet with $V_{rot} = 3$ km/s and $V_{rot} = 6$ km/s with a eastward streaming jet reaching $\pm 20°$ with $V_{Jet} = \mp 6$ km/s and two westward streaming jets at the poles with $V_{WJET} = \pm 2$ km/s (first row) and with $V_{WJET} = \pm 6$ km/s (second row). Note, that the velocity amplitude decreases with $\sim \sin\theta$ (thus the atmospheric velocities at the polar regions are $< V_{WJET}$). The planetary rotation slows down the wind velocities from the westward streaming jet at the poles (i.e. the absolute $V_{LOS}$ decreases) and increases the wind velocities for the eastward streaming jet (i.e. the absolute $V_{LOS}$ increases). See caption in Figure 5 for further information.

Investigating the first row, the line profile for the slower rotating planet looks very similar to the line profiles presented in Section 4.2, where only a eastward streaming jet is considered. This is not surprising, as the Na-lines from the polar regions are only weakly Doppler-shifted by velocities $< V_{LOS} = \pm 1$ km/s, being in the same order to that would be caused by the planetary rotation at such latitudes. In case of the faster rotating planet, the weak Doppler-shift of the absorption lines introduce a stronger contribution at the line core region and less on the line wings, leading to a deeper line core but shallower line wings compared to the profiles shown in Section 4.2, where westward streaming jets are absent.

Investigating westward streaming jets with larger velocity amplitude (second row), the line profile for the slower rotating planet does not show a significant difference compared to the weaker westward streaming jet case (see flux difference in the third row). The line profile becomes the same as shown in the first row Section 4.2, as the line of sight velocity becomes $< V_{LOS} = \pm 3$ km/s, which is the same absolute velocity amplitude compared to a planet with $V_{rot} = 3$ km/s, but only streaming to the opposite direction. Due to the symmetry, this results in a scenario where westward streaming jets can not be distinguished from the effect of planetary rotation without time dependent line profile analysis. For the faster rotating planet, even more interesting, the line of sight velocity at the poles becomes $V_{LOS} = V_{WJET} + V_{rot} = 0$, thus leads to zero Doppler-shift of the absorption lines, so that a stagnant atmosphere is mimicked at the polar regions. The absorption lines from these regions increase the contribution in the line core, so that the line profiles look deeper and narrower compared a planet with the same planetary rotation but absence of such streaming jets.

The required spectral S/N values show that the faster $V_{WJET}$ cases are slightly easier to detect than the slower ones for both type of rotating Jupiter type planets, but the differentiation of the profiles with different $V_{WJET}$ requires way more S/N, especially for the slower rotating planet.

In conclusion, the combination of eastward and westward streaming zonal jets can show line profiles which are deeper and narrower compared to line profiles broadened only by planetary rotation. This effect increases on the one side for decreasing width of the eastward streaming jet and on the other side increasing the contribution from Na- lines with little Doppler-shift from the polar regions. Note, that increasing the westward streaming jet velocity, larger Doppler-shifts will be the case vanishing out this effect, making this result very case dependent and only arising for $V_{LOS} = V_{WJET} + V_{rot} \approx 0$.

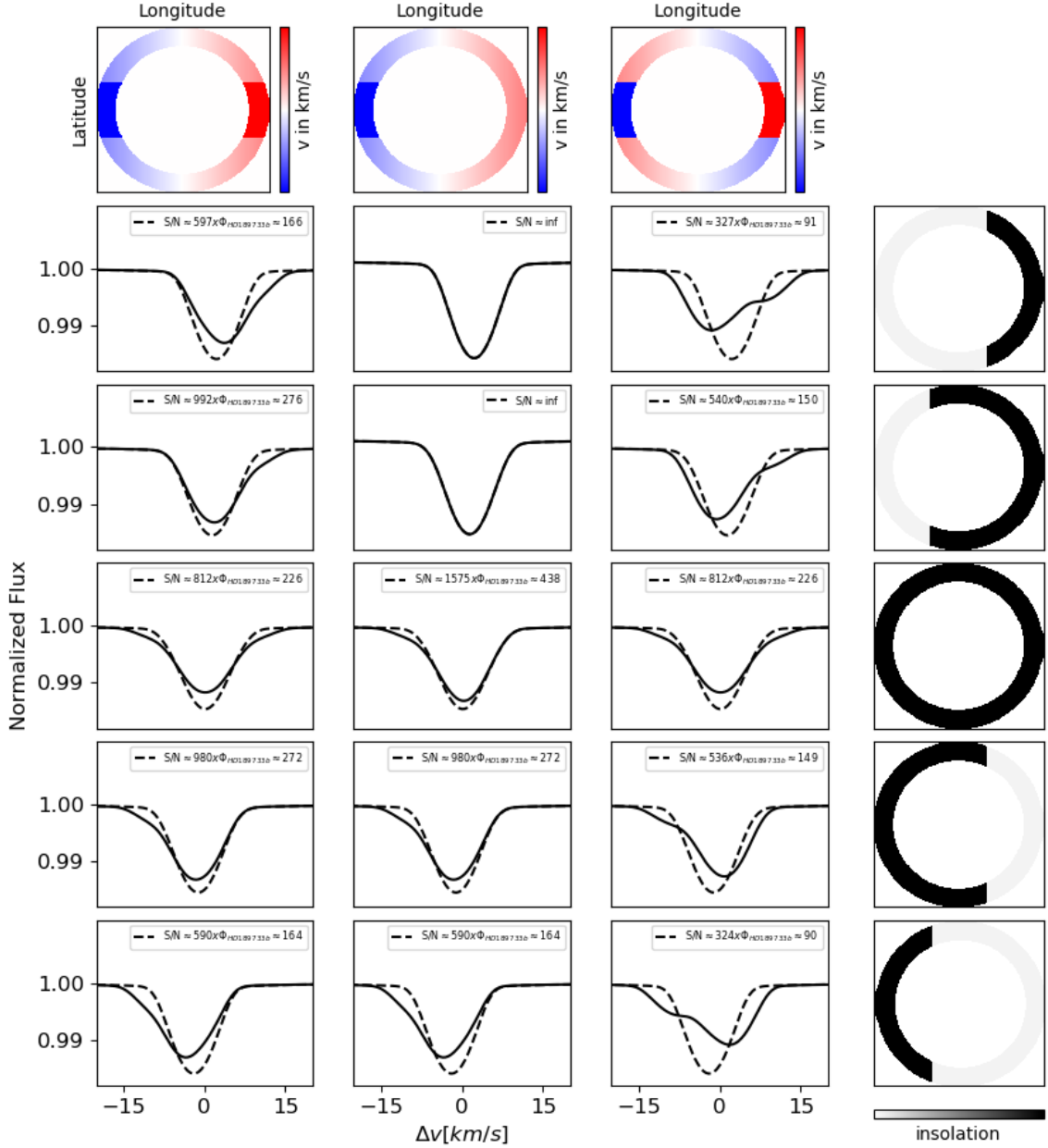

**Figure 6.** Time dependent line profiles for different snapshots during ingress and egress for a Jupiter type planet with $V_{rot}$ = 3 km/s. Column 1, 2 and 3 show the same scenarios as presented in Section 4.2, Section 4.3 and Section 4.4, respectively. The dashed lines consider only the planetary rotation for comparison. The fourth column shows the different snap shots during ingress and egress, whereas the black shaded region illustrates the illuminated atmospheric region. The label shows the $3\sigma$ spectral mean S/N value with $\Phi_{HD189733b} \approx 0.279$. Each column shares the x-axis and the panel with the line profiles share the y-axis.

### 4.5 Time dependent line profile variation during ingress and egress

The previous scenarios were focused on the resolved absorption signature during the second and third contact. But different broadening mechanisms (e.g. expansion, hydrodynamic, thermal and the ones presented in this work) can be degenerate and thus not possible to discern them in this case (Cauley et al. 2020). As the line profiles can look very similar for different wind patterns only accounting for the in-transit phase (as demonstrated), line profiles inferred during ingress and egress can be used with the aim to distinguish between the different wind patterns on the exoplanets (Cauley et al. 2020). For instance, Louden & Wheatley (2015) resolved spatially winds on HD189733b investigating the Na-lines during ingress and egress showing different excess velocities on the trailing and leading limb.

To investigate on timely resolved profiles, Figure 6 shows a Jupiter type planet rotating with $V_{rot}$ = 3 km/s and an eastward streaming jet into the same direction with $V_{Jet}$ = ∓6 km/s reaching a latitude of 20° (first column, see also Section 4.2), the same scenario with a third line of sight velocity on the leading limb (second column, see also Section 4.3) and an eastward streaming jet reaching a latitude of 20° with westward streaming jets on the poles with $V_{WJet}$ = ± 6 km/s (third column, see also Section 4.4). The fourth column shows the illuminated part of the planet at which the line profiles are investigated.

**Table 1.** Na-D2 line properties found for different exoplanets.

| Exoplanet | HD189733b | WASP-52b | WASP-49b | WASP-127b | WASP-17b | WASP-76b | KELT-20b |
|---|---|---|---|---|---|---|---|
| $T_{\text{equilibrium}}$ [K] | 1191 | 1315 | 1399 | 1400 | 1755 | 2228 | 2261 |
| FWHM [Å] | [1]0.52 ± 0.08 [8]0.64 ± 0.04 | [2]0.424 ± 0.036 | [3]0.42 ± 0.1 | [4]0.367 ± 0.097 | [5]0.20 ± 0.08 | [6]0.619 ± 0.174 [4]0.400 ± 0.065 | [7]0.17 ± 0.04 |
| Line contrast [%] | [1]0.64 ± 0.07 [8]0.72 ± 0.05 | [2]1.31 ± 0.13 | [3]1.99 ± 0.49 | [4]1.144 ± 0.270 | [5]1.3 ± 0.6 | [6]0.373 ± 0.091 [4]0.57 ± 0.08 | [7]0.32 ± 0.05 |

Note: ,[1]Wyttenbach et al. (2015), [2]Chen et al. (2020), [3]Wyttenbach et al. (2017), [4]Žák et al. (2019), [5]Khalafinejad et al. (2018), [6]Seidel et al. (2019), [7]Casasayas-Barris et al. (2019), [8]Casasayas-Barris et al. (2017). As for WASP-17b only the Gaussian width was shown, the FWHM here is derived via FWHM = $\sigma \times 2.355$. For KELT-20b, the LC and FWHM is the mean value from the HARPS-N and CARMENES data shown. The FWHM were shown in velocity units and is transformed here to Å.

Investigating the first column with an eastward streaming jet, the line deformation over time is symmetric. The line profile shows a redshift induced from the leading limb during ingress and a blueshift induced from the trailing limb during egress. Furthermore, the streaming jet affects the line wings due to the enhanced contribution from Doppler-shifted Na- absorption at ingress and egress. This is evident through the comparison to the dashed line profile, where only the planetary rotation is considered.

Adding an asymmetry to the wind pattern by decreasing the line of sight velocity on the leading limb by a factor of three (second column), the line profile also becomes redshifted during ingress. This profile is not distinguishable from a planet which only shows planetary rotation, as the line of sight velocities for both cases become very similar. On the other side, the line profile on the trailing limb during egress is identical with the line profile in the first column, thus showing a blueshift with affected line wings. However, since the line profiles are different at the leading limb for the symmetric and asymmetric wind pattern scenarios, the scenarios can be still distinguished.

The strongest line profile deformation is observed when eastern and western streaming jets within the planetary atmosphere are included (third column). In this case, the line profiles arising from poles and equatorial region are shifted to opposite directions for the same snapshot, showing much broader and shallower lines, which are not symmetrical due to the different velocity contributions. The line deformation is symmetric over time for the ingress and egress phase, but opposite to the other scenarios, a blueshift is suggested on the leading limb during ingress which becomes a redshift on the trailing limb during egress. Interestingly, this scenario is not distinguishable from the scenario shown in the first column, if one accounts only for the line deformation during the second and third contact (see also Section 4.4), showing the importance of accounting for line deformation during the ingress and egress phase.

The required spectral S/N values for ingress and egress investigations are larger compared to the previous Scenarios, as only snapshots can be used and not several co-added spectra during the second and third contact. Note, that also the S/N for the fourth row is calculated as a snapshot for better comparison, although here in-transit spectra could be co-added. The westward polar jet stream scenario required the lowest S/N values during ingress and egress, followed by the conventional zonal jet stream reaching latitudes of ±20°. For the asymmetric jet stream velocity case, only the egress snapshots deviate from the dashed line profiles.

Note, that including day-to-nightside winds would introduce additional blueshift to all line profiles shown here. This would increase the blueshift and decrease the redshift of the line profiles shown during ingress and egress phase.

## 5 DISCUSSION

### 5.1 Detectability of atmospheric wind pattern

The investigation of different wind patterns in Section 4 showed that S/N values in the order of few dozens to few hundreds per resolved spectral resolution element in the stellar spectrum are needed to detect the wind pattern significantly. Current high-resolution spectrographs at large telescopes such as PEPSI (Potsdam Echelle Polarimetric and Spectroscopic Instrument, see e.g. Strassmeier et al. (2018))) or ESPRESSO (Echelle SPectrograph for Rocky Exoplanets and Stable Spectroscopic Observations, see e.g. Pepe et al. (2010)) might be able to reach such values within few transit observations.

For comparison, the star HD189733 was observed with PEPSI (14.10.2017) at a exposure time of 30 min and a spectral resolution of R ≈130 000. Extrapolating the S/N per spectral resolution element for an exposure time of 4 min results in an S/N value of ∼160 at ingress/egress and around ∼130 to ∼40 within the second and third contact inspecting the stellar Na-line (not shown here). The S/N values for the different orbital phases are different as during the transit the atmospheric absorption becomes Doppler-shift due to the orbital motion of the planet around the stellar Na-line core. The values show that such wind patterns are detectable with current high-resolution spectrographs within few transits. But although high-resolution spectrographs could be able to differentiate between the different wind patterns, there is a degeneracy especially with other effects which could introduce line broadening e.g. vertical upward winds (Seidel et al. 2020) and atmospheric turbulences (Keles et al. 2020), which is discussed in Section 5.3.

### 5.2 Investigating the resolved Na-D2-lines on different exoplanets

The study of gaseous exoplanet atmospheres revealed the strong absorption of Na- lines. To investigate if those lines show line broadening, which could be induced by winds, Na-D2 lines inferred from high resolution transit observations of different exoplanets will be analyzed. Table 1 shows a list of resolved Na-D2 lines where a line contrast (LC) and FWHM from a Gaussian fit is stated (see the caption for the corresponding authors). The planets are ordered by their equilibrium temperature inferred from the [2]TEPCat database (Southworth 2011). There is a tendency that the FWHM is decreasing for planets with a larger equilibrium temperature. However, these planets have different properties which could lead to different absorption profiles, which need to be accounted for.

---

[2] https://www.astro.keele.ac.uk/jkt/tepcat/

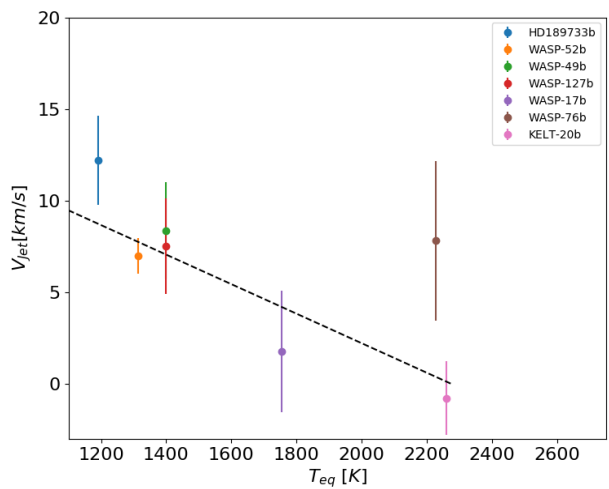

**Figure 7.** Inferred broadening values for the Na-D2 lines detected on different exoplanets presented in Table 1. The y-axis shows the excess velocity i.e. $V_{Jet} = V_{bro} - V_{rot}$.

To investigate on this, the *petitRADTRANS* code (Mollière et al. 2019, 2020) is employed to calculate synthetic high resolution transmission spectra for all planets shown in Table 1. For this, an isothermal temperature profile and different reference pressure level but the same solar Na abundance and a mean molecular atmospheric weight of 2.33 is assumed. The synthetic spectra are converted via $(1-(R_{pl}/R_{star})^2)$ and normalized making them comparable to the observational spectra. The stellar and planetary parameters are taken from the TEPCat database.

It is important to recall that transit observations probe the ~mbar pressure regimes at the terminator, where the temperature profile may not be isothermal, but appropriate for a first order approximation. As the temperature at the terminator is not precisely known and differs from the equilibrium temperature, synthetic lines with different isothermal temperatures were computed in steps of 200 K ranging from 1000K to 9000K and their corresponding FWHM determined. Assuming that a possible line broadening is due to a superrotating zonal jet stream where one can consider flux conservation, the equivalent width of the Gaussian profiles in Table 1 is calculated and used to determine the LC of a Gaussian profile which has the same FWHM as the synthetic line profile. This procedure is applied to enable a more appropriate comparison of the line profiles as the alkali lines show wide Lorentzian line wings compared to the Gaussian line profiles. Due to this, the normalization of such spectra is usually a crucial step affecting the line depth. But as the Lorentzian lines have very sharp line cores, the determined FWHM do not change for slight changes of the line depth. Applying a sanity check by introducing 30% larger line depths showed no significant effects on the results.

The line profile is broadened (using the model presented in Section 2, assuming $(R_{pl} < R_{Atmo} < 1.3 \times R_{pl})$) in steps of 0.2 km/s and compared (±FWHM around the line core) with the observational profiles from Table 1, until a minimum $\chi^2$-value is determined. This approach yield very similar broadening values for the different synthetic profiles with different isothermal temperature, showing that the change in FWHM of the spectral lines for the different temperature is small compared to the broadening value required to match the observations. The error bars are calculated using the uncertainties stated for the line properties in Table 1 by determining the broadening for the Gaussian profiles varied by ± 1-$\sigma$ in LC and FWHM. Figure A1 shows the comparison between the observational line profiles from Table 1 and the line profiles inferred in this work.

Figure 7 shows the broadening values $V_{bro}$ for the Na-D2 lines subtracted from their expected planetary rotational velocity $V_{rot}$. The excess velocity which is left over is attributed here to a superrotating atmosphere, which is assumed to arise from a zonal jet stream (i.e. $V_{Jet} = V_{bro} - V_{rot}$). For HD189733b and WASP-76b the mean $V_{Jet}$ values are shown inferred for the different investigations. There is a indication that the exoplanets with lower equilibrium temperature tend to show less broadening of their atmospheric Na-D2 lines.

As demonstrated, such line broadening can be explained by zonal jet streams on the planetary surface. If the broadening is induced by such a streaming jet, the correlation would imply that HJ type planets favor more the existence of zonal jet streams compared to UHJ type planets. One explanation for this could be that the thermal ionization of alkali lines may induce drag to winds by Lorentz forces (Perna et al. 2010; Batygin & Stevenson 2010; Zhang et al. 2018), which is stronger on UHJ. Increasing the amount of ionized particles, the Lorentz drag and the inflation of Jupiter type planets due to ohmic dissipation (where kinetic energy is converted into heat which becomes deposit in the interior) increases (Zhang et al. 2018). Thorngren & Fortney (2018) showed that this inflation of the giant planet atmospheres peaks at $T_{eq}$ = 1500K, decreasing for larger temperature. The result shown here strengthens this idea that magnetic drag of ionized particles is more efficient for planets with $T_{eq}$ > 1500K, decreasing the efficiency of atmospheric circulation and thus jet stream velocities. Except for WASP-76b, all planets with $T_{eq}$ > 1500K show jet-velocities which are consistent with zero within their error bar, matching well into the big picture. Note, that WASP-76b has quite large error bar and that quite different Na-D2 line properties were measured for this exoplanet. Atmospheres with moderate temperature and high alkali amounts may favor the existence of strong zonal streaming jets which can yield stronger broadened absorption lines, whereas hotter planets may experience thermal ionization of those alkalis reducing the wind strengths of zonal jets, leading to less broadened line profiles.

A result strengthening this picture is shown by Zhang et al. (2018), where the authors showed that the phase offset of the "hot spot" which is shifted by winds is stronger for giant planets with lower irradiation temperature (therefore hinting on stronger jet streams), however, increasing again for irradiation temperatures > ~3400 K (see their Figure 14). Note that the heat recirculation efficiency can increase on UHJ type planets with very high $T_{eq}$ due to the dissociation and recombination of $H_2$ (Bell & Cowan 2018), as shown for instance for KELT-9b (Wong et al. 2020). Another way around, such drags lead to inefficient heat redistribution within the atmospheres. This increases the day- and nightside temperature contrast and triggers stronger day-to-nightside winds on UHJ type planets, which result in stronger blueshifts of spectral lines in transmission spectra (Ehrenreich et al. 2020).

Applying a linear fit to the data (black dashed line) in Figure 7 using the Python package *scipy.optimize.curvefit* (Jones et al. 2001), the following correlation between the equilibrium temperature and excess velocity is inferred:

$$V_{Jet} = -0.0081^{\pm 0.0021} \times T[eq] + 18.33^{\pm 3.17} \quad (2)$$

The errors correspond to the 1-$\sigma$ standard deviation error, showing a ~3.9-$\sigma$ deviation compared to a straight line, indicating that the correlation is significant.

Although the existence of zonal jet streams is an established theory, where the findings in Figure 7 match into the big picture, the velocity amplitudes are quite large and very strong wind speeds should be considered cautiously (see also Section 5.3). The approach demonstrated has large error sources to consider. The employed model yields the broadening values under the assumption that the planetary rotational axis is perpendicular to the orbital plane. For the exoplanets where this is not the case, the velocity amplitudes will be underestimated. Although all targets are observed using high resolution instruments, the difference in the instrumental setup and data reduction technique could introduce significant deviations in the stated line profile properties in Table 1, even for absorption signature detected on the same exoplanets (Wyttenbach et al. 2020). This is also suggested in Table 1, where different authors find different line properties for the Na-lines resolved for the same exoplanet. Moreover, using high resolution observations in which one mainly probes the line cores of these absorption lines, where Gaussian approximation may a valid assumption, the Gaussian profiles do not fit perfectly to the observational Na-lines and will introduce large errors. For instance, Keles et al. (2020) inferred broadening values around ∼10-12 km/s for the Na-D2 line for the exoplanet HD189733b (see their Figure 5), which would lead to $V_{Jet} \approx 7$-9 km/s, a lower value than the broadening value inferred here using the Gaussian line properties. Therefore, the inferred broadening values from this work should be considered cautiously, as the error bars are most probably underestimated. However, this work shows only a first order approximation for the correlation between the broadening of the Na-D2 lines for higher equilibrium temperature (which is also suggested only by inspecting Table 1) and it does not aim to determine the atmospheric wind pattern, which would be out of the scope of this work.

The line broadening of the Na-D2 lines could be explained also by another mechanism instead of a jet stream. For example, such a strong line broadening was not inferred for the water lines (Brogi et al. 2016) or the potassium line (Keles et al. 2020) on the exoplanet HD189733b, which one may expect. However, these lines probably probe lower altitudes compared to the Na-lines, which may explain the difference in line broadening. Contradicting to this, with decreasing pressure, the day-to-night temperature contrast increases (Zhang et al. 2018), thus probably slowing down zonal wind velocities. More advanced modelling effort would be needed to prove this, which is out-of the scope of this work. Future observations resolving the Na-lines on different exoplanets will help to prove this prediction.

### 5.3 Caveats

The results presented in this work depend strongly on the assumptions made, which are discussed here.

This work considers only the broadening of spectral lines due to velocity broadening, assuming uniform wind pattern for the different scenarios i.e. that the wind velocities only change with the line of sight contribution. But winds can vary also strongly as a function of speed and height (Brown 2001) leading to non-uniform wind pattern e.g. depending on the frictional drag within the atmosphere (see Figure 4 in Miller-Ricci Kempton & Rauscher (2012)), resulting possibly in more bimodal line profiles (Showman et al. 2012). Furthermore, the assumption of flux conservation i.e. that the equivalent width of the absorption lines is conserved, is only valid if wind velocities vary only with latitude and longitude, but not valid for wind variations within the line of sight chord (Brown 2001), which would affect the wind velocities inferred in this work.

A more accurate way of introducing broadening to the line profiles could be to introduce the broadening to the height dependent opacity contributions instead of assuming equal opacity at each depth (see e.g. Brown (2001)) and using 2D (see e.g. Cauley et al. (2020)) or even 3D modelling approaches (see e.g. Flowers et al. (2019)).

Another assumption made in this work is that the day-to-nightside wind acts the same everywhere on the terminator region, leading to a net blueshift of the spectral lines without affecting the line widths. However, day-to-nightside winds are expected to stream at a lower pressure level compared to where zonal jet streams arise (see e.g. Showman et al. (2008)), where both winds could combine and show a more complex non-uniform wind pattern. Comparing Figure 6 from Showman et al. (2012), day-to-nightside winds may differ also for different latitudes and longitudes on the terminator region, which would affect the line profile deformation even more. However, the investigation of this phenomenon considering both effects and including the pressure dependence is out of the scope of this work.

Note, that also other kinds of atmospheric motions could introduce a large velocity signature into line profiles, such as evaporative winds caused by the absorption of high energetic radiation (Debrecht et al. 2020) or planetary magnetic fields which may propel ions to high velocities at higher altitudes where they recombine at the same speed (Seidel et al. 2020). Moreover, vertical upward winds are of special interest, which are not considered in this work. Seidel et al. (2020) investigated the effect of different kind of winds on the Na-line profiles on HD189733b, including superrotating atmospheres, vertical upward winds as well as the combination of both. The authors showed that the Na-line broadening on HD189733b requires the additional effect of vertical upward winds because zonal winds alone were not able to reproduce the observed Na-line profiles. Such vertical upward winds are perpendicular to the planetary surface and result in a small line of sight contribution within transmission spectra, leading to large wind velocities, which would be required for enhanced line broadening. But winds in form of zonal jet streams can combine with such vertical wind components, introducing a degeneracy, complicating the determination of zonal jet streams and their properties. In conclusion, line broadening by winds may not be attributed only to zonal jet streams.

The combined velocities at the photospheres of exoplanets can reach 100% or larger velocity values than the equatorial rotational velocities (Spiegel et al. 2007). The largest velocities in the scenarios presented in Section 4 can reach ∼12 km/s, being similar to the velocities inferred from the investigation of the observational Na-D2-lines (see Figure 7). From the observational point of view, velocities close to those values are inferred in different investigations. As mentioned, Keles et al. (2020) showed that the Na- lines on HD189733b need to be broaden around 10 km/s to match the observational line profiles. Recently, Ehrenreich et al. (2020) investigated the nightside condensation of iron on the UHJ WASP-76b inferring the absorption of atmospheric iron on the trailing limb, which shows a blueshift of -11 ± 0.7 km/s. Even more recently, Cauley et al. (2020) investigated the Balmer-lines and broadening by day-to-nightside winds, jet streams, thermal and rotation broadening on WASP-33b inferring a rotational velocity of around 10.1 ± ∼0.9 km/s.

From the theoretical point of view, although drag-free simulations provide wind velocities up to ∼15 km/s (Miller-Ricci Kempton & Rauscher 2012), super-sonic winds below the exobase are not possible and introducing drag (which is a more realistic assumption) can decrease significantly the maximum possible wind velocities, showing a slight tension with the observational results.

The sound speed of non-degenerate molecular hydrogen is approximately $V = 2.4 \times (T/1000 \text{ K})^{0.5}$ km/s (Goodman 2009). For moderate temperatures, this velocity should be around 3-4 km/s in the upper atmospheric layers (Snellen et al. 2008). Thus, the inferred large wind speeds may be overestimated in this work. However, note that stellar gravity and magnetic field could move the sonic point to lower altitudes, enabling larger wind velocities at higher altitudes (Wyttenbach et al. 2020). But the purpose of this work is to show the qualitative effect of super-rotating streaming jets on spectral line profiles. Modeling approaches accounting for non-uniform wind patterns as well as frictional drag within the atmosphere is out of the scope and purpose of this work.

## 6 SUMMARY

Close-in exoplanets experience strong stellar insolation. As these planets are expected to be tidally locked, the constant energetic bombardment on the dayside builds up a so called "hot spot" and triggers atmospheric winds due to the uneven heating of the hemispheres and planetary rotation (Brown 2001). Such winds are mainly categorized into either day-to-nightside winds at higher altitudes which push the atmosphere into the line of sight of the observer during transit or zonal jet streams at lower altitudes which rotate into the direction of planetary rotation (Miller-Ricci Kempton & Rauscher 2012; Showman et al. 2012). The main effect of day-to-nightside winds are blueshifts of absorption signature within the planetary atmosphere, whereas zonal jet streams tend to broaden these absorption lines.

This work shows the effect of zonal jet streams in high resolution transmission spectra on Jupiter type planets rotating with $V_{rot}$ = 3 km/s and $V_{rot}$ = 6 km/s by introducing rotational broadening to synthetic line profiles. Different scenarios were presented showing different wind patterns with eastward and westward streaming zonal jet streams which reach different latitudes and introduce deformation to absorption lines. This work shows that especially during the ingress and egress phase the line deformation can help to distinguish different wind patterns within the atmospheres.

In different exoplanet atmospheres, the Na lines were resolved showing large differences in line widths in comparison to what would be expected only by accounting for planetary rotation. Comparing the observational findings with modeled line profiles, this work demonstrates that planets with larger equilibrium temperatures show less broadened Na-D2 absorption lines. This shows that strong zonal jets arise more efficiently on HJ compared to UHJ consistent with theoretical expectations (Showman et al. 2012), if the broadening is attributed to zonal jets. One explanation could be that the ionization of alkali atoms introduces drag into these winds in atmospheres of UHJ (Perna et al. 2010), leading to inefficient heat redistribution and slowing down the wind velocities. Note, that the recirculation efficiency can increase for very large atmospheric temperature due to the dissociation and recombination of $H_2$ (Bell & Cowan 2018), leading to zonal jet streams also on UHJ. However, one should note that the explanations given in this work should be adopted cautiously due to the large error sources that were discussed.

Atmospheric dynamics can induce further to planetary rotation line broadening and shifts on the absorption lines, depending on the wind profiles of the probed atmospheric region (Flowers et al. 2019). Evidently, especially the alkali lines can be a good tracer for such winds on exoplanets.

## ACKNOWLEDGEMENTS

I am deeply grateful for the scientific discussion and support by Matthias Mallonn, Xanthippi Alexoudi, Laura Ketzer, Daniel Kitzmann and Thorsten A. Carroll. I also thank the anonymous referee for the helpful suggestions increasing the quality of the paper.

## DATA AVAILABILITY

The data underlying this article will be shared on reasonable request to the corresponding author.

## APPENDIX A: APPENDIX

This paper has been typeset from a T<sub>E</sub>X/LAT<sub>E</sub>X file prepared by the author.

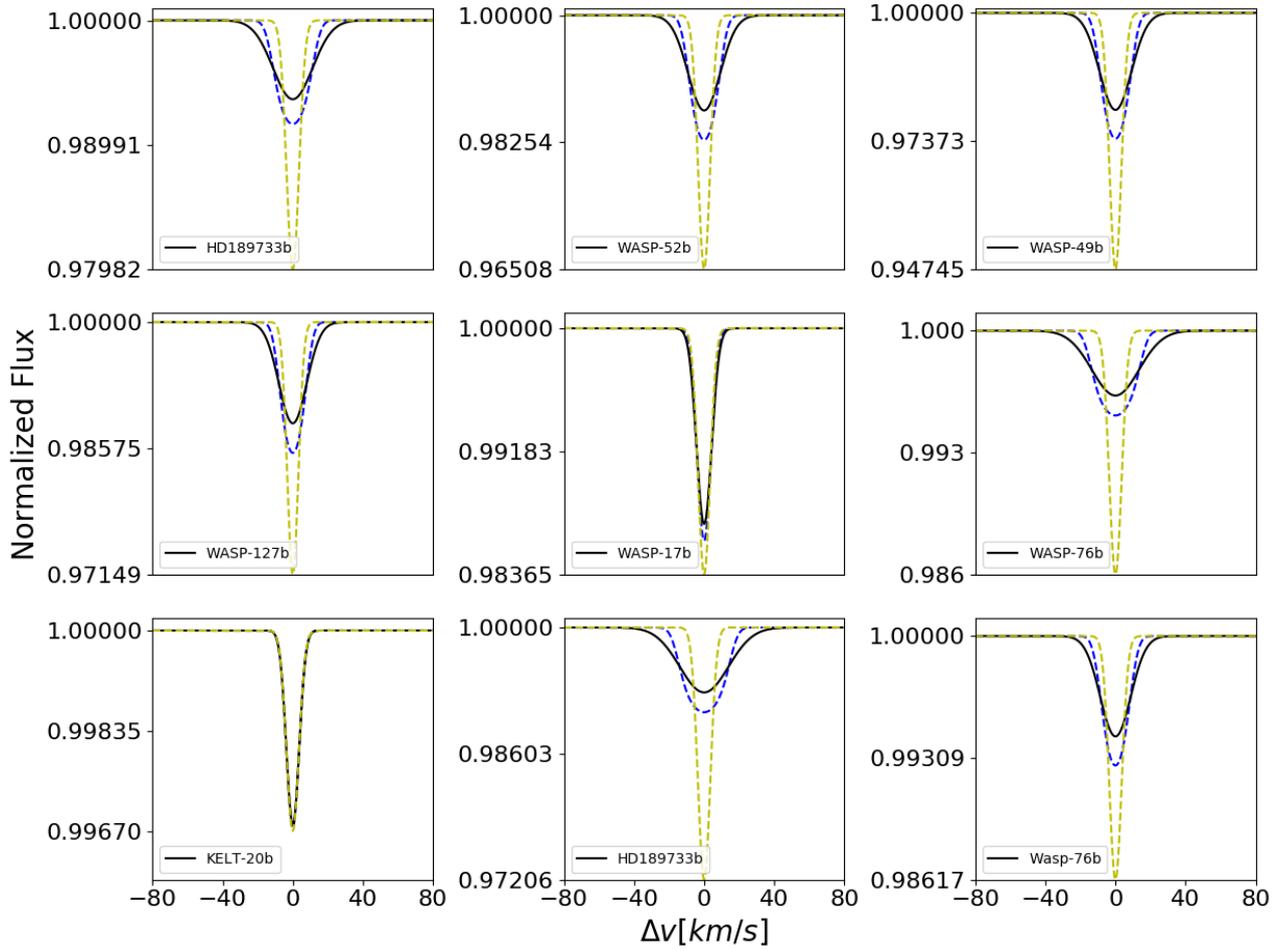

**Figure A1.** Shown are the best matching line profiles between the observational (see also Table 1 for comparison) Gaussian line profiles (black solid) and the broadened line profiles inferred from the equivalent width comparison (blue dashed). The dashed yellow line shows the unbroadened line profile. The different columns share the x-axis.